%% file: main.tex
\documentclass[sn-basic,centre]{sn-jnl}

\usepackage{graphicx}%
\usepackage{multirow}%
\usepackage{amsmath,amssymb,amsfonts}%
\usepackage{amsthm}%
\usepackage{mathrsfs}%
\usepackage[title]{appendix}%
\usepackage{xcolor}%
\usepackage{textcomp}%
\usepackage{manyfoot}%
\usepackage{booktabs}%
\usepackage{algorithm}%
\usepackage{algorithmicx}%
\usepackage{algpseudocode}%
\usepackage{listings}%

\usepackage{fontawesome5}
\usepackage{comment}
\usepackage{makecell}
\usepackage{multirow}
\usepackage{enumitem}
\usepackage{array}
\usepackage[final,commandnameprefix=always,defaultcolor=red]{changes} 

\theoremstyle{thmstyleone}%
\theoremstyle{thmstyletwo}%

\theoremstyle{thmstylethree}%

\begin{document}

\title[Animated Transition between Node-Link and Parallel Coordinates Visualizations]{Animated Transition between Node-Link and Parallel Coordinates Visualizations}


\author*[1]{\fnm{Abdulhaq Adetunji} \sur{Salako}}\email{abdulhaq.salako@uni-rostock.de}

\author[1]{\fnm{Hannes} \sur{Hagen}}\email{hannes.hagen@uni-rostock.de}

\author[1]{\fnm{Christian} \sur{Tominski}}\email{christian.tominski@uni-rostock.de}

\affil*[1]{\orgdiv{Institute for Visual \& Analytic Computing}, \orgname{University of Rostock}, \orgaddress{\street{Albert-Einstein-Straße 22}, \city{Rostock}, \postcode{18059}, \country{Germany}}}


\abstract{Multi-faceted data visualization typically involves several dedicated views. To create a comprehensive understanding of the data, users have to \chreplaced{mentally}{manually} integrate the information from the different views. This integration is hindered by context switches between views \chadded{and usually requires interactive methods such as brushing and linking.} Animated transitions have also been shown to be able to mediate context switches and improve understanding. Yet, most existing animated transitions consider only basic views showing the same data facet.
In this work, we study how the gap between node-link diagrams, showing graph structure, and parallel coordinates plots, showing multivariate attributes, can be narrowed via smooth animated transitions. Based on two design goals (traceability and swiftness), we outline a partial design space including several design options. These inform the implementation of two alternative transition variants: a basic variant with plain interpolation and an advanced variant that uses our design space and accepted animation techniques, including staging and staggering. In a preliminary study, we asked seven participants for qualitative feedback. We found that the swiftness of the basic variant is preferred, while the traceability of data items is better with the slower advanced variant.}




\maketitle

\input{chapter/introduction}

\input{chapter/related_work}

\input{chapter/smooth_transitions}

\input{chapter/study}

\input{chapter/discussion}

\input{chapter/conclusion}

\backmatter

\bmhead{Acknowledgments}
\chadded{We thank the anonymous JapanVis reviewers for their valuable comments and helpful suggestions.} We also gratefully acknowledge funding support for the project \emph{iV-Morph: Interactive Visual Metamorphosis for Multi-view Data Exploration} from the DFG under grant number \href{https://gepris.dfg.de/gepris/projekt/514630063?language=en}{514630063}.

\bibliography{references}

\end{document}

%% file: chapter/introduction.tex
\section{Introduction}
\label{sec1}

The visual analysis of complex data typically requires multiple visual representations, each visualizing a different data facet or showing the data from a different perspective~\citep{Tominski21FlexibleVA}. For example, when analyzing multivariate graphs, one may need representations that show the graph structure, the attributes of nodes and edges, the temporal evolution of the graph and its attributes, and the graph's spatial frame of reference~\citep{Kerren2013IntroductionTM, Hadlak15Multifaceted}.

Gaining insight from multiple views requires users to integrate the depicted information into a comprehensive picture of the data. This integration process is hindered by context switches that occur when users shift their attention from one visual representation to the other~\citep{Baldonado2000GuidelinesFU}. Context switches between substantially different visual representations tend to be particularly costly. In addition to interactive highlighting and visual linking~\citep{Collins07VisualLinking,Steinberger11VisualLinks}, animated transitions have been proven useful to mediate the context switches between different visual representations~\citep{Heer07Animation, Fisher10Animation, Ruchikachorn15Learning}. Yet, existing animated transitions are typically limited to basic chart types, such as bar chart and pie chart. Despite recent research on understanding the design space of animations~\citep{Thompson20Animated, Kim21Gemini, Zong23AnimatedVega} there is still relatively little known about how to design animated transitions for non-trivial visualization techniques showing different data facets.

In this work, we address this gap by studying the design of animated transitions between node-link diagrams (NL) and parallel coordinates plots (PC), which has not yet been considered in the literature. NL is widely used for showing the topological structure of graphs, whereas PC typically visualize multivariate data. When used in combination, both visualizations can facilitate the analysis of complex multivariate graphs~\citep{Shannon2008MultivariateGD, Guo09FlowMapping, Kerren2013IntroductionTM}, which makes the two techniques interesting candidates to be mediated with animated transitions. \chadded{Transitions between NL and PC are significant in helping a range of users, from experts analyzing complex graphs to learners seeking to understand data relationship.}A particular challenge of designing animated transitions between NL and PC lies in the fact that they focus on different data facets, NL on graph structures and PC on multivariate attributes. 

As a first step toward animated NL-PC transitions, we explore alternative transition designs that aim to satisfy transition swiftness and transition traceability as two key design goals. \chadded{Our exploratory research yields a partial view on a larger design space for NL-PC transitions. In addition to new NL-PC-specific design considerations,} \chreplaced{established}{Different} animation techniques, including staging \citep{Heer07Animation} and staggering \citep{staggeringChevalier14}, will be applied to make the animated transitions comprehensible to arrive at an overall smooth context mediation. We tested our initial transition designs in a preliminary qualitative user study. According to the user feedback, our transitions are comprehensible and can help in tracking changes when switching views. \chdeleted{The study results suggest that the staging method and the staggering method can improve traceability, although they may reduce the swiftness of NL–PC transitions.}

%% file: chapter/related_work.tex
\section{Related Work}
\label{sec2}

Our work is motivated by the challenges in visualizing multi-faceted data, of which multivariate graphs are a prominent example. The solution we present in this work further builds on and is inspired by previous research on animated transitions in visualization.

\subsection{Visualizing Multi-Faceted Data}

Multi-faceted data comprise several data facets that require dedicated techniques for their appropriate visual representation~\citep{Hadlak15Multifaceted}. Time ($T$), space ($S$), multivariate attributes ($A$), and structural relationships ($R$) are data facets commonly found in multi-faceted data~\citep{Tominski20IVDA}. How to visualize these data facets individually has been extensively studied (see \cite{Aigner23VisTimeOrientedData} for $T$, \cite{Kraak20Cartography} for $S$, \cite{Ward15DataVis} for $A$, and \cite{Tamassia13GraphVis} for $R$). However, the combined visualization of multiple data facets, i.e., a multi-faceted visualization, remains challenging~\citep{Filipov23NetworkVis}.

A common solution is to use multiple visualization views~\citep{Baldonado2000GuidelinesFU}, each representing a selected primary data facet (e.g., NL for $R$ or PC for $A$) and optionally integrating secondary or tertiary data facets at a (much) lower level of detail~\citep{Tominski21FlexibleVA}. \cite{Javed2012ExploringTD} explored ways to combine multiple views to form so-called composite visualization views, including view arrangements and embeddings. Side-by-side views provide additional details but can disrupt the analysis due to cognitive gaps between views, making it harder for users to integrate information. Embedded views address this by co-locating views in the same space to reduce cognitive load and improve comprehension. However, effective integration requires careful visual encoding to prevent clutter.

\cite{Tominski21FlexibleVA} introduced the idea of flexible visual analytics (FVA) as a solution to both cognitive and visual gaps in multivariate data exploration. FVA aims to mitigate these challenges by fluid transitions between user-relevant views, allowing users to seamlessly navigate between different perspectives while maintaining contextual awareness. We build on the idea of FVA and put a particular focus on transitions between NL and PC for seamless integration of structural relationships $R$ and multivariate attributes $A$.

\subsection{Animated Transitions}

Many early studies on animation were inspired by principles from traditional cartoon animation \citep{Chang1993AnimationFC} and were motivated by applications such as explaining temporal data \citep{Schell2007SocioeconomicDO} and depicting state transitions \citep{Tang2020NarrativeTI, Bludau2021UnfoldingEF, Berger2023ComparingNO}. Animations have both pros and cons in visualization \citep{Tversky02Animation, Fisher10Animation}. On the one hand, they can enhance user engagement and orientation. On the other hand, \chadded{animations consume time, and} poorly designed or unnecessary animations can clutter information displays, making them difficult to perceive accurately and hindering comprehension. \chadded{Different design strategies exist to improve animations, including staging \citep{Heer07Animation}, where specific changes occur in distinct phases, and staggering \citep{staggeringChevalier14}, which introduces small delays between individual object movements.}

\cite{Fisher10Animation} categorized six types of animations for visualization. Our work falls under his category of \textit{changing the representation} by smoothly transitioning between visualizations to enhance data comprehension. This has proven effective in several studies. For example, \cite{Heer07Animation} transitioned scatterplots into bar charts and found evidence that animated transitions can help preserve users' mental maps. The NodeTrix technique by \cite{Riche2007NodeTrixHR} uses animated transitions for partial transformations between node-link and matrix representations. Previous work by \cite{Yuan2009ScatteringPI} has also studied transitions involving parallel coordinates and scatter plots. \cite{Ruchikachorn15Learning} found that animated morphing of visual representations helps in teaching visualization. Interestingly, some previous work on animated transitions involve not two but three visualizations to improve the overall design. For example, \cite{Yang2020TiltMI} transition from choropleth map via prism map to bar chart, and \cite{Kiesel2023SmoothTB} use a polycurve star plot as an intermediate representation during the transition between parallel coordinates and scatter plots.

Although, the design space of animations in visualization has been thoroughly explored \citep{Thompson20Animated, Kim21Gemini, Zong23AnimatedVega}, existing research mainly focuses on transitions between views that share the same data facet \chadded{\citep[e.g.,][]{Huth23Smallscale}}. Transitions between views showing different data facets (of the same data) are rare \chadded{\citep[e.g.,][]{Yang2020TiltMI}}. Our work aims to narrow this gap and investigates transitions between relationship-oriented NL and attribute-oriented PC views.

%% file: chapter/smooth_transitions.tex
\section{Smooth Transitions between NL and PC}
\label{sec3}

To ensure smooth, interpretable, and meaningful animated transitions, we first need to define design goals. Based on these goals, we discuss different design options and propose concrete designs for animated transitions between NL and PC.

\subsection{Design Goals}

Building on previous work on animated transitions, we identified the following two key goals for our animated NL-PC transitions. 

\begin{itemize}[label={}]
    \item \faLightbulb{} \textbf{Traceability:} In order to function as a mediator, animated transitions should be traceable. The traceability goal aims for transitions that are easy to follow and help users transfer their mental model between views.

    \item \faRunning{} \textbf{Swiftness:} Animated transitions require time. The swiftness goal aims to run transitions at a reasonable speed to keep the time costs low, but not too fast to hinder comprehension.
\end{itemize}

While \faLightbulb{} \chreplaced{emphasizes reducing mental effort by making transitions easy to follow}{focuses on human resources, emphasizing the ease of following transitions}, \faRunning{} focuses on \chreplaced{time efficiency }{ time resources}, ensuring transitions happen quickly. \chreplaced{This introduces }{There is} a natural tension between \faLightbulb{} and \faRunning{}. A too rapid transition may compromise the ability to track elements effectively. Contrarily, making each and every detail fully traceable surely slows down the overall data analysis process.

The design of animated transitions that balance \faLightbulb{} and \faRunning{} is influenced by various factors. Engagement: A transition should keep users engaged. Short-term memory: A transition should not take too long, as users might find it difficult to remember what they saw. Uniformity: A transition should show uniform and consistent visual changes. Correspondence: A clear relationship between data objects and their visual representations should be ensured throughout a transition. Perceptual capacity: A transition should align with the viewer's perceptual capacity. Validity: Only accurate and meaningful intermediate representations should be displayed during a transition to avoid data misinterpretation.

\subsection{Representation Mapping}

In order to design an animated transition between two visual representations, one first has to understand how the two representations are constructed individually and then define a representation mapping \chreplaced{that specifies how data items and graphical primitives in one visual representation are mapped to those in the other visual representation. This mapping explains how visual elements showing the same or different data items are connected across the two views.}{The mapping describes how data items and graphical primitives in one visual representation are mapped to graphical primitives showing the same (or other) data items in the other visual representation.} In our case of animated NL-PC transitions, we assume the data to be multivariate graphs where the graph nodes have multivariate attributes.

\paragraph{Characterizing NL and PC Representations}

NL representations visualize entity-relationship data ($R$). Entities (or nodes) are typically represented as dots of certain size and color. Relationships (or edges) are visualized as links between dots. For our animated NL-PC transition, we focus on nodes as data items and dots as graphical primitives.

In PC representations, multivariate attributes ($A$) are the primary data facet. Attributes are represented as parallel axes, and data items (e.g., nodes of a graph) are visualized as polylines intersecting these axes according to their multivariate attribute values. For the animated NL-PC transition, we focus on nodes with attribute values as the data, and polylines and axes as the graphical primitives.

\paragraph{Mapping Between Representations}

The identified data items and graphical primitives are the basis for the mapping between representations. As we use nodes as the data items, dots in NL must be mapped to polylines in PC. Yet, the polylines can be complex when there are many attributes. Therefore, we propose a rather drastic simplification. Instead of mapping a dot to a potentially complex polyline for $n$ axes, we map to a simple line for only $2$ axes. Once this basic dots-to-lines transformation is complete, one can expand from $2$-axes PC to $n$-axes PC using a straightforward accordion transition \citep{slackaccordion2006} \chadded{that stretches out the additional axes so that} \chdeleted{where} the full complexity of the polylines \chreplaced{unfolds}{to unfold} gradually \citep{Bludau25UnfoldablesSTAR}. Additionally, we need to consider that links cease to exist and axes appear during the NL-PC transition. In summary, the following mapping must be animated during the transition:

\begin{itemize}
    \item dots $\rightarrow$ lines
    \item links $\rightarrow$ $\varnothing$
    \item $\varnothing$ $\rightarrow$ axes
\end{itemize}

\subsection{Designing the Animated Transition}

In the following, we present initial design ideas for animated NL-PC transitions. Although, there is a huge design space, we cannot explore it comprehensively in this work. Instead, our goal is to develop first practical solutions to be tested in a preliminary user study. 

\paragraph{General Transition Procedure}

\begin{figure}[h]
    \centering
    \includegraphics[width=1\linewidth]{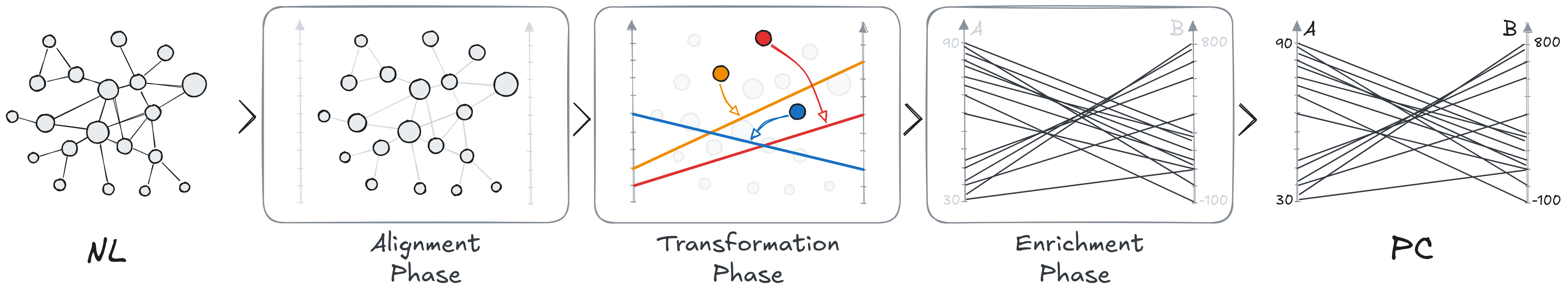}
    \caption{Schematic representation of a general transition procedure consisting of alignment, transformation, and enrichment phases, here illustrated for the transition from a node-link diagram (NL) to a two-axes parallel coordinates plot (PC).}
    \label{fig:procedure}
\end{figure}

The representation mapping suggests that a general 3-phase transition procedure is reasonable consisting of the alignment phase, the transformation phase, and the enrichment phase, as shown in \autoref{fig:procedure}. During the alignment phase, a common ground for the transformation is established, and the view is reduced to the core information that actually gets transformed. In our NL-PC case, \chadded{assuming centered views,} NL links fade-out, the (two) PC axes fade-in, and the dots are maintained. In the transformation phase, the core information is transformed between the two representations, which in our case means that the NL dots become PC lines. This transformation deserves special attention, and hence will be described in more detail later. Finally, in the enrichment phase, details such as axis labels, UI elements, etc. can be added.

\paragraph{Design Considerations for the Transformation Phase}

The design of the transformation phase involves decisions on how to align the spatial arrangement and how to transform the graphical primitives. In our case, we need to convert 2D dots into 1D lines and rearrange the structure-driven NL layout into the attribute-driven PC layout. During the transformation, the viewer will perceive several graphical changes:

\begin{itemize}
    \item Shape ($C_{shape}$): Dots transform into lines.
    \item Size ($C_{size}$): Graphical primitives will change in size (compact dots vs. long lines).
    \item Position ($C_{pos}$): Dots will shift their positions as they become lines.
\end{itemize}

\begin{figure}
    \centering
    \includegraphics[width=1\linewidth]{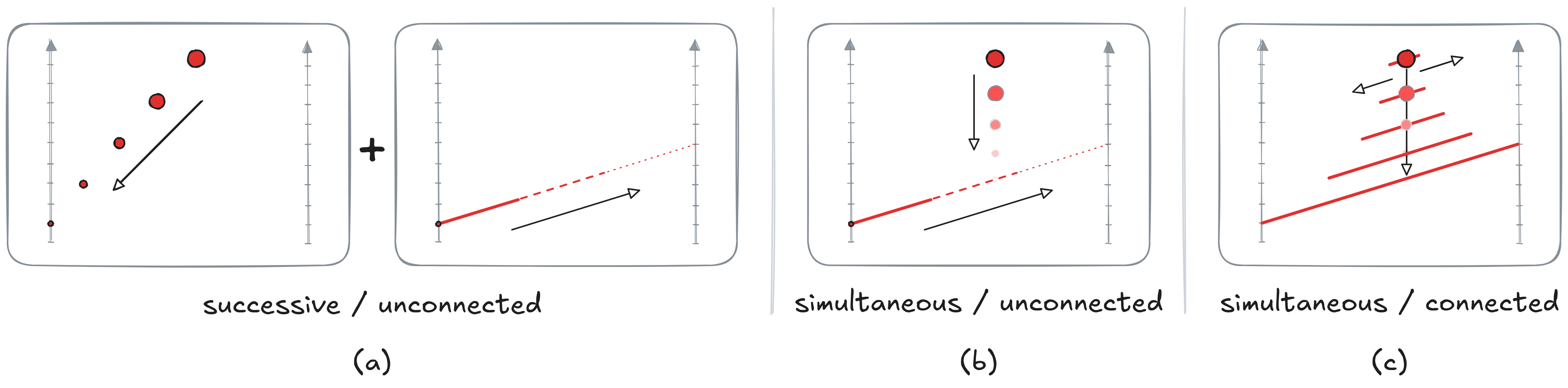}
    \caption{Concrete transition designs showing different transformation approaches.}
    \label{fig:t-designs}
\end{figure}

The question for the transition design is where and when these changes should happen. One can easily imagine a plethora of concrete designs for these steps. Guided by the principles of congruence and apprehension \cite{Tversky02Animation}, we explored different transition strategies, three of which are depicted in \autoref{fig:t-designs}.

A reasonable strategy is to first reposition the dots ($C_{pos}$) to align with their attribute values on the first PC axis. As the dots move, they gradually shrink to small points ($C_{size}$). Once the small points reach their position on the first axis, they elongate into lines ($C_{size}$) that connect to the corresponding attribute value on the second PC axis to complete the transformation. In \autoref{fig:t-designs} (a), one can see that the changes occur \emph{successively} and that dots and lines are \emph{unconnected} \chadded{at the start of the animation}. The successive changes may be good for \faLightbulb{}, but less useful for \faRunning{}. Moreover, the fact that dots and lines are \chadded{initially} unconnected may negatively impact \faLightbulb{}.

As another strategy, illustrated in \autoref{fig:t-designs} (b), one might think of letting lines emanate (simultaneous $C_{shape, size}$) directly from the left axis toward the right axis while simultaneously relocating and shrinking the dots (simultaneous $C_{pos, size}$) so that the lines pass through their corresponding dots \citep[similar to][]{Yuan2009ScatteringPI}. While the \emph{simultaneous} changes are good for \faRunning{}, dots and lines are still \emph{unconnected}, which arguably is disadvantageous for \faLightbulb{}.

Therefore, we explored a third strategy in \autoref{fig:t-designs} (c), where \emph{simultaneous} and \emph{connected} changes happen. Similarly to the previous strategy, dots are relocated and shrink while lines grow \emph{simultaneously} ($C_{pos, size, shape}$). Yet now, two line segments emanate directly from the dots and grow symmetrically toward the data-induced end points on the two axes. With this, dots and lines are \emph{connected} throughout the transition, which may be beneficial for \faLightbulb{}. The simultaneous changes benefit swiftness \faRunning{}, but could also negatively impact \faLightbulb{}.

Based on this initial exploration, we investigated the transformation phase in more detail. In particular, we looked at the shape and size changes $C_{shape}$ and $C_{size}$, which mutually depend on each other quite strongly, and at position changes $C_{pos}$.

\begin{description}
    \item[Designing for $C_{shape}$ and $C_{size}$] The question is how to design the shape and size transition so that viewers can anticipate the transition outcome already early on. A trivial geometric interpolation between a dot and a line is easy to implement, but \chreplaced{certainly sub-optimal as the intermediate shapes would cause unnecessary occlusion.}{is not guaranteed to avoid degenerated intermediate shapes.}
    Therefore, the outlined strategies already suggested shrinking dots while growing lines. The growing line can already be oriented according to the final line. This allows viewers to make an early estimation about the relation ($<, \sim, >$) between the two attribute values ultimately visualized by the line. If anticipating the actual values (rather than their relation) is preferred, one can use a bent line with two arms that always point toward the final endpoints of the line, which represent the attribute values on the axes.

    \item[Designing for $C_{pos}$] The transition between NL layout and PC arrangement requires relocating graphical primitives. To make the relocation easily traceable, one option is to restrict it to 1D \chadded{vertical} movements\chdeleted{along the vertical y-axis}. Alternatively, one can relocate along the shortest path between dot and line. Again, both options have pros and cons. The shortest path reduces the overall length of relocation movements, but it could also lead to confusion due to multi-directional movements. The 1D-only option avoids such confusion, but might lead to overall longer movement paths.
\end{description}

\paragraph{Animation Timing}

On top of the individual design options for $C_{shape}$, $C_{size}$, and $C_{pos}$, there is the need to decide on a timing for them. In principle, changes can take place successively (e.g., $C_{shape} \rightarrow C_{size} \rightarrow C_{pos}$), simultaneously (e.g., $C_{shape, size, pos}$), or a combination of both (e.g., $C_{shape, size} \rightarrow C_{pos}$). Successive changes tend to benefit \faLightbulb{}, whereas simultaneous changes are good for \faRunning{}. Secondly, order and timing can relate to all or to individual data items. This is reflected in established animation techniques, namely \emph{staging} and \emph{staggering}.

\textbf{Staging} Instead of performing all changes simultaneously, staging allows changes to happen one after the other~\citep{Heer07Animation, StagingCrnovrsanin21}. This approach can help establish a logical flow and improve transition clarity. We outline three possible staging sequences based on perceived graphical primitives: 

\begin{itemize}  
    \item \textbf{\(C_{pos} \rightarrow C_{shape} \rightarrow C_{size}\)}:  
    This sequence first applies position change. Once repositioned, dots are converted into line segments, which are then extended in length to form complete lines across the two axes in PC.  

    \item \textbf{\(C_{shape} \rightarrow C_{size} \rightarrow C_{pos}\)}:  
    The transition begins with a shape change, where dots transform into small lines, which then fully extend to complete lines across the two axes in PC. Finally, position adjustments are applied.

    \item \textbf{\(C_{shape} \rightarrow C_{pos} \rightarrow C_{size}\)}:
    The transformation starts by changing dots into small lines, which then move to their correct positions. Once positioned, the lines elongate to span the two axes in PC.
    
\end{itemize} 

\textbf{Staggering} While staging gives structure to the changes that happen during the transition, staggering is concerned with timing the transitions of individual elements~\citep{staggeringChevalier14}. Rather than animating all elements at once, staggering introduces a delay in each element's transformation, so that individual element transitions start one after the other. This does not mean that a previous transition has to complete before another one can start, but only the starts are delayed and several elements can still transition in parallel.

Sorting mechanisms are used to define the order for the staggering. The sorting can be based on attribute values (e.g., data of interest first), spatial position (e.g., smallest position change), or group relationships (e.g., clusters transition together).

\begin{figure}
    \centering
    \includegraphics[width=1\linewidth]{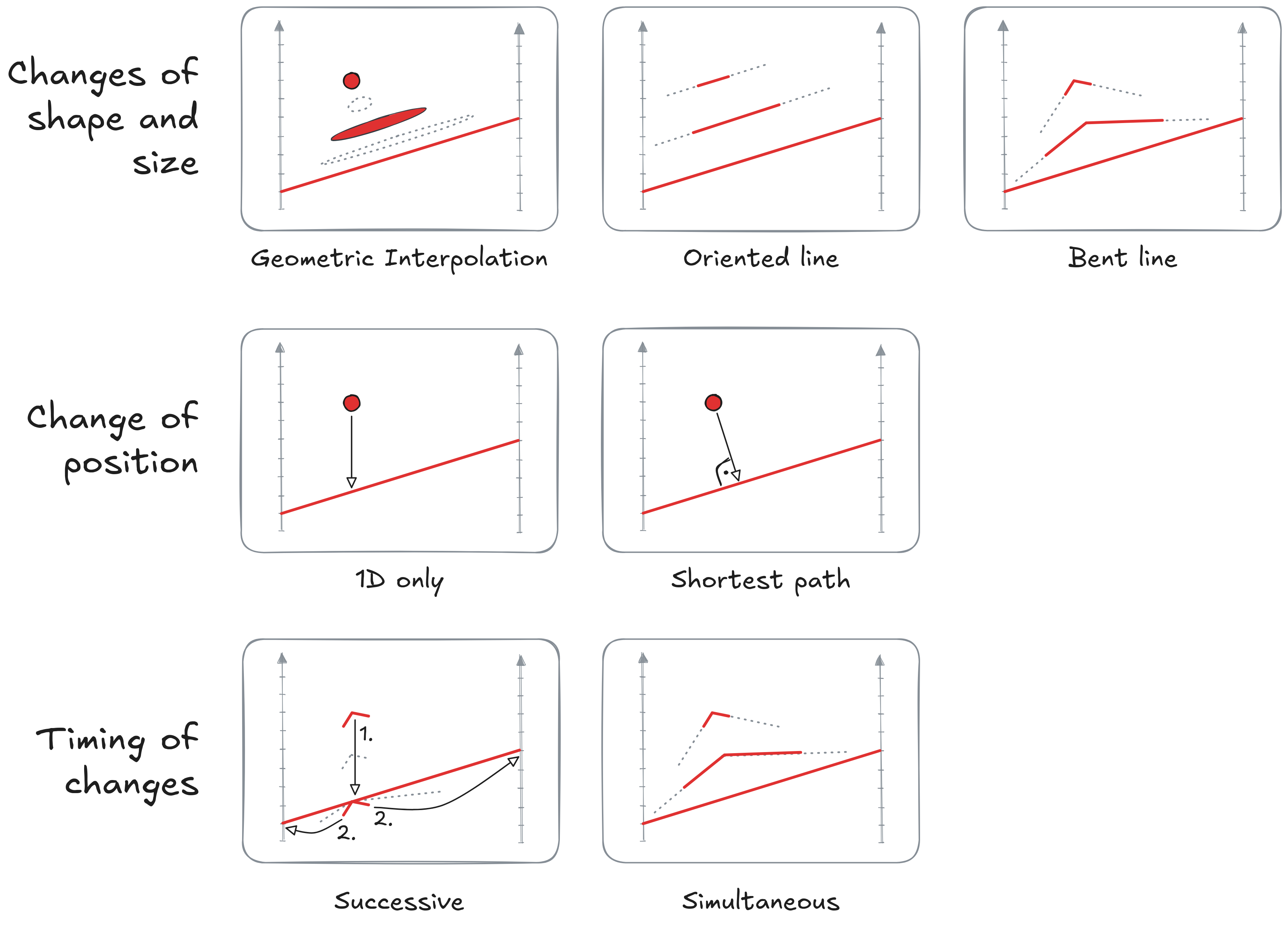}
    \caption{Different design options for transforming a NL dot into a PC line.}
    \label{fig:design-options}
\end{figure}

The different design options for our animated NL-PC transitions are summarized in \autoref{fig:design-options}. They will later inform the implementation of concrete transition variants for the user study.

\paragraph{Completing the transition}

So far, our focus was on designing the transition from NL to PC with only two axes, which \chreplaced{is not the typical use case for PC.}{are not true PC in the literal sense.} The final step to complete the transition would be to expand from $2$-axes PC to $n$-axes PC. We consider this to be doable with a straightforward accordion transformation where further axes unfold \chreplaced{perpendicularly to the PC axes.}{to the right of the second PC axis.} Such accordion transitions have previously been used in the visualization literature \citep{slackaccordion2006,Bludau25UnfoldablesSTAR}.

\medskip

The outlined strategies and design options suggest that there is a huge design space that we could explore only partially in this work. And even our limited design space gives rise to many different animated NL-PC transitions.

\subsection{Implementing the Animated Transition}

Our implementation effort focuses on two different transition variants to be tested in a preliminary user study (see Section \ref{sec:study}) to acquire feedback with respect to the design goals of swiftness \faRunning{} and traceability \faLightbulb{}.

\paragraph{Basic Transition Variant}

In this variant, all changes happen simultaneously and are implemented using trivial geometric dot-to-line shape interpolation. While this makes the transition swift (\faRunning{}$+$), there is no guiding structure, and the intermediate shapes \chadded{are not informative and} might \chreplaced{cause occlusions}{be degenerated}, which may hurt traceability (\faLightbulb{}$-$).

\begin{figure}
    \centering
    \includegraphics[width=1\linewidth]{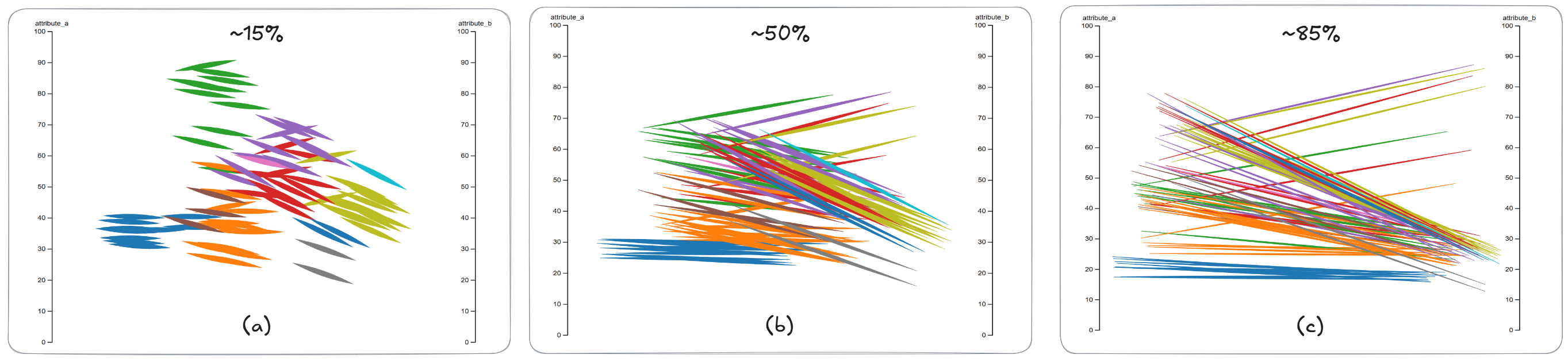}
    \caption{Transformation process from NL to PC illustrating the basic variant where shape, size, and position changes happen simultaneously.}
    \label{fig:basic-variant}
\end{figure}

\autoref{fig:basic-variant} shows a concrete example of such a basic transformation. Figure (a) is early at $\sim15\%$ of the transition, where $C_{pos}$, $C_{shape}$, and $C_{size}$ start happening simultaneously. The figure (b) represents the synchronized progression of all changes about $50\%$ through the transition. Clearly, the trivial geometric interpolation leads to sub-optimal intermediate representations. Finally, figure (c) illustrates the nearly complete transformation at $\sim85\%$, where the dots have almost fully transitioned into lines, completing the transformation phase.

\paragraph{Advanced Transition Variant}

The second variant uses the detailed design options for $C_{pos}$, $C_{shape}$, and $C_{size}$ discussed earlier, including oriented line and bent line as well as 1D-only and shortest path. For the timing, we incorporated staging and staggering to reduce excessive simultaneous movement and makes the transformation easier to follow (\faLightbulb{}$+$). On the other hand, staging and staggering require additional time, which makes this variant less swift (\faRunning{}$-$).

\begin{figure}[b!]
    \centering
    \includegraphics[width=1\linewidth]{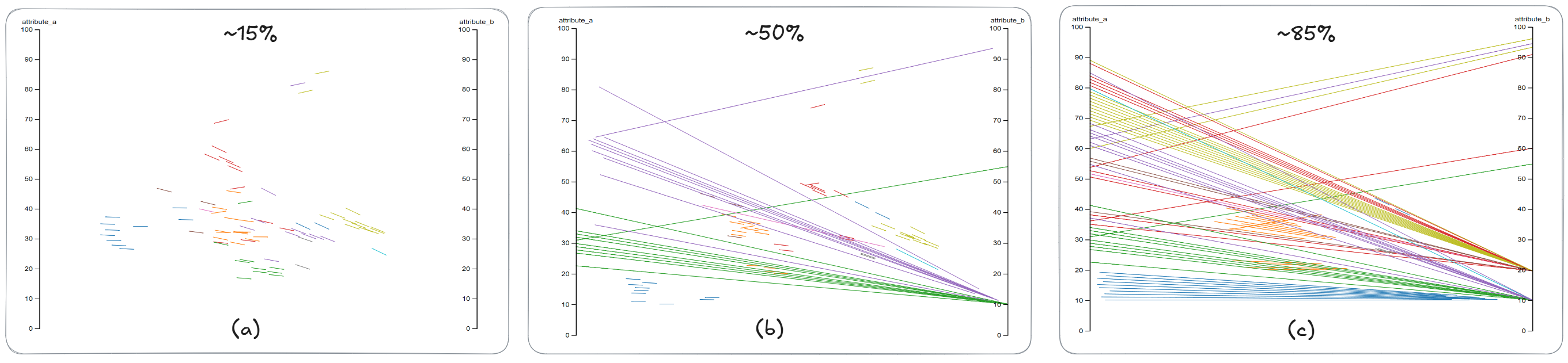}
    \caption{Transformation process from NL to PC using the advanced variant, where staging and staggering are applied to the changes in shape, size, and position.}
    \label{fig:advance-variant}
\end{figure}

\autoref{fig:advance-variant} illustrates an example of staged and staggering transition using the advanced variant. In figure (a), at approximately $\sim15\%$ into the transition, dots have already been transformed gradually into oriented lines $C_{shape}$. Shortly after, $C_{pos}$ starts as the small oriented lines move toward their respective line position in PC. Figure (b) is mid-stage at around $\sim60\%$, where $C_{pos}$ is nearly complete as the last lines reach their position. Then they start extending into full lines. Finally, figure (c) shows the late stage of the transition at $\sim85\%$, where $C_{size}$ is almost done and all lines are nearly fully extended across both axes completing the transformation phase.

\medskip

In this section, we proposed \chreplaced{novel animation designs}{a general procedure} for transitioning between \chadded{NL and PC} views, which show different data facets, and outlined various transitions strategies and design options, partially spanning a design space for animated NL-PC transitions. 

We introduced two concrete transition implementations. The first variant uses a rather trivial strategy that makes the transition swift, but not necessarily easily traceable. The second variant uses staging and staggering to improve traceability, but these require more time making the transition less swift. A video of both variants is provided as \href{https://nextcloud.informatik.uni-rostock.de/s/cCyZGJz5jBA4R5H}{supplementary material}. Next, we report on a preliminary user study that tested the two transition variants.

%% file: chapter/study.tex
\section{Preliminary Study}
\label{sec:study}

It is beyond the scope of this work to fully evaluate even the partial design space of animated NL-PC transitions. Instead, we focus on soliciting qualitative user feedback on the two implemented transition variants and the tension between \faRunning{} and \faLightbulb{}. Such early user feedback can help us optimize the transition design in future work.


\paragraph{Participants}

Our study involved seven doctoral students from a university, 2 female and 5 male, aged between 24 and 30 years. 
Their academic backgrounds included visualization, databases, theoretical computer science, and data security. The majority of the participants (6 out of 7) were not experts in animated transitions or data visualization.

\paragraph{Stimuli, Tasks, and Devices}

The study used the Les Misérables graph~\citep{Koblenz2013}, a widely used dataset in visualization. The fully connected graph has 77 nodes and 254 undirected edges and includes 11 node clusters. Since the graph comes without node attributes, synthetic attributes were added to create a multivariate graph. The data values were generated based on predefined visual PC patterns, such as negative correlation or outliers~\citep{Tominski20IVDA}. 

\begin{figure}[t]
    \centering
    \includegraphics[width=1\linewidth]{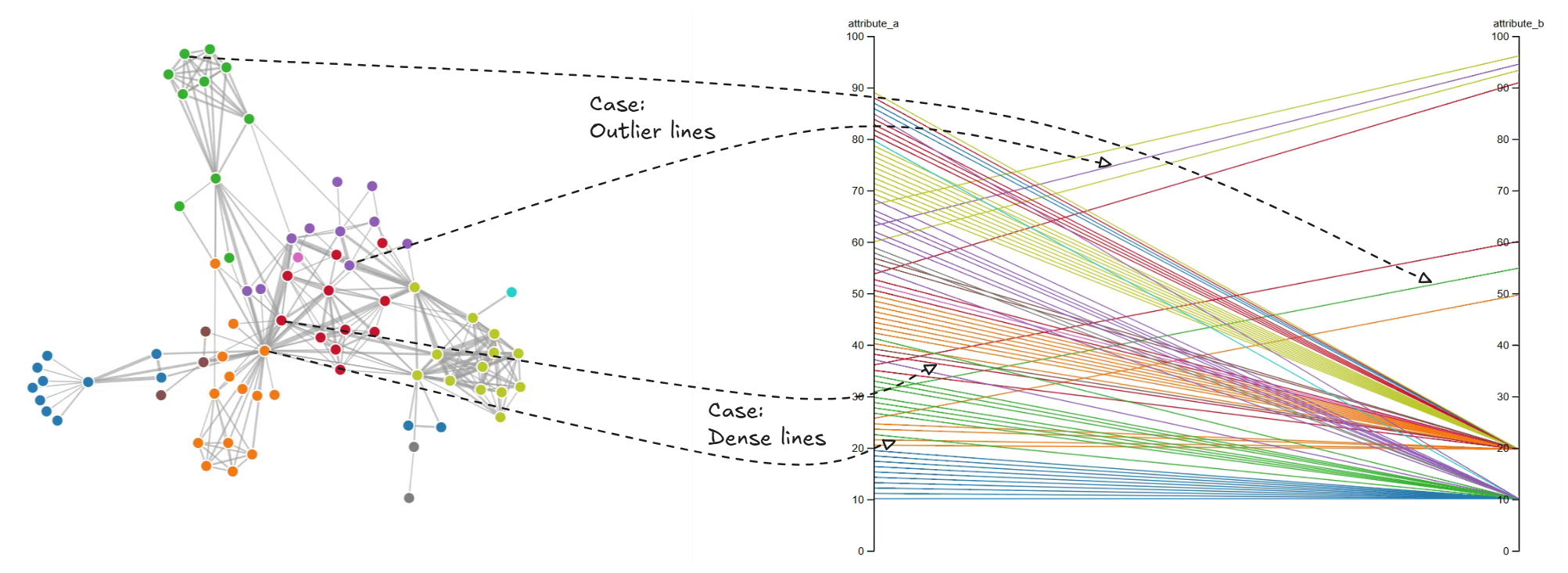}
    \caption{Different cases of tracking tasks.}
    \label{fig:nl-pc}
\end{figure}

The data were visualized as NL and PC as shown in \autoref{fig:nl-pc}, where color indicates cluster affiliation. The two transition variants introduced earlier were used in the study. One variant is a basic geometric animation ($V_{basic}$) and the other is an advanced animation ($V_{adv}$) using our identified design options. In both variants, a 1-second alignment phase was applied before the transformation. In $V_{basic}$, the transformation phase lasted 2 seconds during which all changes ($C_{shape, size, pos}$) happen simultaneously. In $V_{adv}$, the changes happened in 3 stages ($C_{shape} \rightarrow C_{pos} \rightarrow C_{size}$), each stage lasted 2 seconds. The staggering introduced a 0.02-second delay per node and an additional 0.4-second delay per node cluster. As our focus was on the actual transformation, the final enrichment phase was not animated (0 seconds).

The study consisted of tracking tasks where participants had to trace a data element from NL to PC. At the start of a task, a dot in NL was highlighted with a flashing border, which disappeared before the transformation began. After the animated NL-PC transition, participants had to identify the line in PC that corresponds to the dot in NL by clicking the line, after which the correct line flashed for verification. We created different cases (see~\autoref{fig:nl-pc}), where the final lines are rather isolated outliers (easier tracking) or are in rather dense PC regions (more difficult tracking). As we were interested in qualitative feedback only, no time or errors were quantitatively measured.


The study was conducted on the participants' desktop computers, ensuring familiarity with their environment. Each participant used a Full-HD display (1920x1080) with a screen diagonal of 25" to 27". To ensure that participants relied solely on their visual tracking, mouse interaction, particularly the mouse cursor, was disabled during the animated transitions.

\paragraph{Procedure}

Each study session took place with one study participant within a 30-minute time frame. The sessions began with an information block, explaining the motivation for the work and the basics of the NL and PC representations. The two variants were presented, 4 participants started with $V_{basic}$, 3 participants started with $V_{adv}$. For each variant, there were a familiarization, a task, and a questionnaire phase.

First, the participants could familiarize with the transition variant by viewing it several times. Then, participants entered into the task phase, during which two tracking tasks (outlier and dense) needed to be performed. The elements to be tracked were varied to prevent participants from recognizing previously tracked elements.
After the two tracking task, a questionnaire was conducted where the questions from Table~\ref{tab:questionnaire} had to be answered using a five-level Likert scale, except for last questions where a transition phase had to be identified.

\begin{table}[t]
    \caption{Questionnaire for rating our animated NL-PC transitions.}
    \label{tab:questionnaire}
    \begin{tabular}{@{}lp{10cm}@{}}
        \toprule
        \textbf{Aspect} & \textbf{Questions} \\
        \midrule
        \multicolumn{2}{l}{\textit{General animation properties}} \\
        \midrule
        Complexity I & Is the transformation demanding? \\
        Complexity II & Is the number of elements transforming simultaneously demanding? \\
        Speed & Is the speed of the transformation appropriate? \\
        Total duration & Is the total duration of the animation appropriate? \\
        Intuitiveness & Is the transformation intuitive? \\
        Clarity & Are the individual changes clearly recognizable? \\
        Continuity & Does the transformation proceed smoothly? \\
        Predictability & Is the transformation predictable? \\
        Staging & Do the intermediate stages help in understanding the animation? \\
        Staggering & Is staggering useful to understand the transformation? \\
        \midrule
        \multicolumn{2}{l}{\textit{Element tracking}} \\
        \midrule
        Traceability I & Could the node be tracked for the outlier case? \\
        Traceability II & Could the node be tracked for the dense case? \\
        Speed & Did the animation speed affect tracking? \\
        Density & Do densely crowded lines affect tracking? \\
        Colors & Does the color coding affect tracking? \\
        Problem phase & During which transformation step is tracking the most difficult? \\
        \midrule
    \end{tabular}
\end{table}

\paragraph{Results}

The participant ratings were analyzed, where 5 denoted a highly positive effect (e.g., ``strongly helpful'') and 0 indicated no effect (e.g., ``not helpful''). Per survey question, extreme min and max ratings were removed, and the remaining five ratings were averaged. The results were extracted with regard to two key aspects: general animation properties (first part of Table~\ref{tab:questionnaire}) and element tracking (second part of Table~\ref{tab:questionnaire}).

Table~\ref{tab:study_results1} shows how participants rated the general animation properties for $V_{basic}$ and $V_{adv}$. The results indicate that $V_{basic}$ performed better in continuity (5.0 vs. 3.4) and total duration (3.2 vs. 2.6), suggesting that a more brief transformation can maintain a stable flow. However, $V_{adv}$ outperformed $V_{basic}$ in clarity (2.2 vs. 4.6) and predictability (2.8 vs. 3.8), indicating that staging and staggering helped users understand the transformations more substantially compared to the simultaneous transition in $V_{basic}$. The intermediate stages appearing only in $V_{adv}$, rated 4.6, likely improved comprehension of the transition process. Staggering was rated not particularly helpful (1.0) for supporting the understanding of the transformation.

\begin{table}[t]
    \centering
    \caption{Participant ratings (0-5) for general animation properties.}
    \begin{tabular}{p{4cm}cc}
        \toprule
        \textbf{Aspect} & \textbf{Variant $V_{basic}$} & \textbf{Variant $V_{adv}$} \\
        \midrule
        Complexity I & 3.4 & 3.6 \\
        Complexity II & 2.2 & 3.6 \\
        Speed & 3.4 & 3.2 \\
        Total duration & 3.2 & 2.6 \\
        Intuitiveness & 4.2 & 4.2 \\
        Clarity & 2.2 & 4.6 \\
        Continuity & 5.0 & 3.4 \\
        Predictability & 2.8 & 3.8 \\
        Staging & - & 4.6 \\
        Staggering & - & 1.0 \\
        \toprule
    \end{tabular}
    \label{tab:study_results1}
\end{table}

\begin{table}[t]
    \centering
    \caption{Participant ratings (0-5) for element tracking.}
    \begin{tabular}{p{4cm}cc}
        \toprule
        \textbf{Aspect} & \textbf{Variant $V_{basic}$} & \textbf{Variant $V_{adv}$} \\
        \midrule
        Traceability I (outliers case) & 4.4 & 5.0 \\
        Traceability II (dense case) & 2.2 & 4.4 \\
        Speed & 3.0 & 3.2 \\
        Density & 1.0 & 2.6 \\
        Colors & 4.8 & 4.4 \\
        Problem Phase & - & $C_{pos}$ \\
        \midrule
    \end{tabular}
    \label{tab:study_results2}
\end{table}

The survey results focusing on the element tracking are presented in Table~\ref{tab:study_results2}. The results show that $V_{adv}$ improved traceability, particularly for the case of lines in dense PC regions (2.2 vs. 4.4). For the outlier case, both variants performed well (4.4 vs. 5.0), but traceability was still slightly higher $V_{adv}$. The density impact score (1.0 vs 2.6) suggests that staging and staggering helped a bit mitigate tracking challenges in dense regions. Interestingly, the influence of color was rated consistently (4.8 vs. 4.4), indicating that color plays an important role for traceability for both variants. Only for $V_{adv}$, we asked which change was most difficult to track, to which most participants responded with $C_{pos}$.

Overall, the findings suggest that our animated NL-PC animated transitions work well (for single element tracking). As expected, staging improves traceability \faLightbulb{} at the costs of swiftness \faRunning{}. The impact of staggering was limited. Other aspects, such as coloring and the positioning of elements, seem to have an impact, but this needs to be investigated further in future work.

%% file: chapter/discussion.tex
\section{Discussion}
\label{sec:disc}

\chadded{Our work on NL-PC transitions is still early research and as such has limitations. These primarily pertain to our exploration of the design space and to the conducted user study.}

\paragraph{Design limitations}

\chadded{Our work is mostly focused on animated transitions between NL and PC two axes. Already for this quite narrow view on multi-faceted animated transitions, the design space seems to be huge.}

\chadded{The design space becomes even larger when considering the following aspects, which we could not yet take into account. So far, we assume that the views to be animated are well defined in the sense that a reasonable visual mapping of the data exists (e.g., node attributes mapped to dot size and PC axes). Yet, one could also vary the visual mapping as part of the animated transition design. Similarly, one could consider designing the animated transition with respect to data size, data structure, or data distribution, meaning that the transition would be different if certain characteristics are mapped. Such animations could then emphasize clusters or trends in the data or focus the viewers attention to specifically interesting data elements.}

\chadded{Likewise, one could imagine task-specific animation designs. We only considered the tracking of data elements between views and the identification of data values. Yet, visual analysis tasks are diverse (e.g., analyzing correlations in PC or tracing cost-efficient paths in NL), and it seems promising to consider this diversity to provide animated transitions depending on the user's task at hand. Dealing with the added design complexity and the even larger design space will be a formidable research challenge.}

\paragraph{Study limitations}

\chadded{Our user study could only shed little light on the qualities of animated NL-PC transitions. The number of study participants was limited and we could only test two transition variants. Tracking single nodes was the only tested task, whereas real-world multi-faceted analysis tasks, such as understanding cross-view correlations were not considered. It remains an open question how NL-PC transitions work for different data characteristics. We may hypothesize that data elements in clusters could be easier tracked and understood during an animated transition that data elements that disperse across the entire visualization. Yet, how to deal with such split-attention problems, is subject to further research and testing.}

\chadded{Our study could only test two different timings of the animated transitions, a rather short basic variant and a rather long advance variant. Yet, it would also be interesting to study if and how user-controlled transitions could improve both \faLightbulb{} and \faRunning{}.}

\chadded{With regard to \faLightbulb{}, it would also be relevant to test how previous user experience with NL and PC impacts the understanding of an animated transition. Specifically for \faRunning{}, it would be interesting to test staging and staggering, and their particular timing in more detail. Studying all these relevant and important aspects would make it necessary to conduct more and larger experiments. }


%% file: chapter/conclusion.tex
\section{Conclusion}
\label{sec5}

In this work, we explored the design of animated transitions between NL and PC to bridge the conceptual gap between these visualization techniques. We discussed several design options and outlined a partial design space. Two alternative transition variants were implemented, a basic variant with plain geometric interpolation and an advanced variant using our design space, staging, and staggering. A user study provided preliminary qualitative feedback \chreplaced{indicating that}{that indicates the general suitability of} animated NL-PC transitions \chadded{can be understood reasonably well for simple tasks.} \chreplaced{The study}{and} also confirmed the need to balance traceability \faLightbulb{} and swiftness \faRunning{} carefully.

In future work, we plan to investigate animated transitions further in two directions. 
First, our work on animated NL-PC transitions has by far not explored the design space for animated transitions comprehensively. We implemented and tested only two alternative transition variants. Therefore, the design space of NL-PC transitions needs to be explored further, and the effect of the different design options on user comprehension when switching between NL and PC visualizations should be evaluated more thoroughly.

Second, we studied animated transitions to mediate view switches between NL and PC, which focus on structural relationships ($R$) and multivariate attributes ($A$) as relevant data facets. For true multi-faceted data exploration, animated transitions would be needed for all possible \chadded{combinations and} pairs of data facets, including time ($T$) and space ($S$). For example, how can we transition NL into a \chreplaced{space-time cube}{time line} representation or PC into \chadded{multivariate glyphs on} a geographic map? 

Ultimately, we envision what we call \textit{shape-shifting views} that are capable of transforming themselves into any visualization mediated by smooth transitions. Multiple data facets can then be explored by user-controlled animated shape-shifting that seamlessly takes the user from one data facet to another.